# Ternary Nitride Semiconductors in the Rocksalt Crystal Structure

*Sage R. Bauers*[*,1], *Aaron Holder*[1,2], *Wenhao Sun*[3,4], *Celeste L. Melamed*[1,5], *Rachel Woods-Robinson*[1,3,4], *John Mangum*[5], *John Perkins*[1], *William Tumas*[1], *Brian Gorman*[5], *Adele Tamboli*[1], *Gerbrand Ceder*[3,4], *Stephan Lany*[1], *Andriy Zakutayev*[*,1]

[1]Materials Science Center, National Renewable Energy Laboratory, Golden, Colorado 80401, United States

[2]University of Colorado, Boulder, Colorado 80309, United States

[3]Materials Science Division, Lawrence Berkeley National Laboratory, Berkeley, California 94720, United States

[4]University of California Berkeley, Berkeley, California 94720, United States

[5]Colorado School of Mines, Golden, Colorado 80401, United States

[*]Corresponding authors

## Abstract

Inorganic nitrides with wurtzite crystal structures are well-known semiconductors used in optoelectronic devices. In contrast, rocksalt-based nitrides are known for their metallic and refractory properties. Breaking this dichotomy, here we report on ternary nitride semiconductors with rocksalt crystal structures, remarkable optoelectronic properties, and the general chemical formula $Mg_xTM_{1-x}N$ (TM=Ti, Zr, Hf, Nb). These compounds form over a broad metal composition range and our experiments show that Mg-rich compositions are nondegenerate semiconductors with visible-range optical absorption onsets (1.8-2.1 eV). Lattice parameters are compatible with growth on a variety of substrates, and epitaxially grown $MgZrN_2$ exhibits remarkable electron mobilities approaching 100 $cm^2$ $V^{-1}s^{-1}$. Ab initio calculations reveal that these compounds have disorder-tunable optical properties, large dielectric constants and low carrier effective masses that are insensitive to disorder. Overall, these experimental and theoretical results highlight $Mg_{G-3}TMN_{G-2}$ rocksalts as a new class of semiconductor materials with promising properties for optoelectronic applications.

Nitride materials are relevant to several industrial and technological fields, and historically are separated into two families. The first family is main-group metal nitride semiconductors with wurtzite crystal structure, typified by (Al,Ga,In)N, which are known for direct band gaps and high carrier mobilities and found in various optoelectronic devices.[1,2] Over the last few decades these materials have become particularly important due to proliferation of solid state lighting, radio frequency transistors, and high information density optical storage media.[1–3] The second family is transition metal (*TM*) nitrides with rocksalt structures, such as the transition metal mononitrides TiN, VN, and CrN, which are often found as hard coatings in industrial tools and diffusion barriers in semiconductor devices.[4] In these compounds, the open d-shell of the transition metal leads to metallic behavior; some of them also exhibit superconducting transitions approaching 20 K.[5] While some rocksalt transition metal nitrides (e.g. ScN) are narrow indirect gap semiconductors[6], and metallic wurtzite $ZnMoN_2$ was recently reported[7], these are rare exceptions to the more general trends.

Recently, increased attention has been devoted to semiconducting II-IV-$N_2$ ternaries, which are structurally similar to III-N wurtzite compounds but with the main group $III^{3+}$ metal replaced with group $II^{2+}$ and group $IV^{4+}$ metals (for example, $Zn^{2+}$ and $Ge^{4+}$ instead of $Ga^{3+}$).[8] These heterovalent wurtzite-derived compounds have potential for increased control of electronic properties both by changing the metal composition and tuning the degree of cation disorder.[9,10] It is not clear if a similar approach might be used in transition metal nitrides with rocksalt structures. On one hand, the group IV transition metals usually exist in 3+ valence states in nitrides, making them metallic (e.g. TiN).[11] On the other hand, introduction of electropositive low-valence alkaline earth (*AE*) cations, such as $Mg^{2+}$, could move transition metals into higher oxidation states, closing the d-shell and inducing semiconducting behavior.[12] Even though computational studies suggest that many such rocksalt derivatives can be formed,[13,14] only a few of these compounds (e.g. SrZrN$_2$, SrHfN$_2$) have been synthesized, and reports of thin-film synthesis and functional properties are currently lacking.[15–17]

Here, we report on a family of new ternary nitride semiconductors with rocksalt structures and the general formula Mg$_{G-3}$TMN$_{G-2}$, where G is the group number of transition metal *TM* (*TM* = Ti, Zr, Hf, Nb). First-principles structure predictions reveal rocksalt-derived structures, where multiple cation ordering motifs on the metal sublattice can be close in energy. Each of the compounds, synthesized as thin films via sputtering, exhibits diffraction peaks corresponding to a simple rocksalt structure and lattice parameters compatible with epitaxial growth on a variety of substrates. Mg-rich compositions exhibit semiconducting optoelectronic properties with visible-range optical absorption onsets. MgZrN$_2$ grown on MgO shows electron mobilities approaching 100 cm$^2$ V$^{-1}$s$^{-1}$ at ca. 10$^{18}$ cm$^{-3}$ electron density. Ab initio calculations suggest these compounds have disorder-tunable optical properties, disorder-tolerant electronic properties, low effective masses, and large dielectric constants. Overall, these results broaden the landscape of semiconducting nitrides to ternary materials with rocksalt-derived crystal structures.

First we hypothesize about possible structures found in ternary rocksalt-derived nitrides. Ternary rocksalt-derived nitrides can exist in several extended structures, but can be locally described as edge sharing cation/anion octahedral motifs, just like in the case of oxides.[18] The motifs for N-*AE*$_3^{2+}$*TM*$_3^{4+}$ octahedra corresponding to *AE*$^{2+}$*TM*$^{4+}$N$_2$ stoichiometry are shown in **Figure 1a**. They may form in a clustered arrangement, with like-cations grouped together such that they define an octahedral face, or in a dispersed arrangement, with like-cations separated across the octahedron. Alternatively, in a disordered state the motifs can be mixed, and deviations from the ideal *AE*/*TM* ratio surrounding the N atoms may occur. However, this becomes energetically more costly the further they deviate from local charge neutrality, as recently shown for tetrahedrally coordinated ternary nitrides.[19] Following the observed structural preferences in oxides and similar ternary nitrides, where radius ratio rules have been used to determine the coordination and adopted motif,[13,15,18] $MgTiN_2$ is expected to exhibit clustered octahedral motifs, whereas $MgZrN_2$ and $MgHfN_2$ are expected to have dispersed motifs. Extending the same motif arguments to $AE_2^{2+}TM^{5+}N_3$ stoichiometries, $Mg_2NbN_3$ is expected to have a clustered N-$AE_4^{2+}TM_2^{5+}$ motif. Interestingly, within the confines of this framework, local charge neutrality could not be achieved with octahedrally coordinated G=6 metals, suggesting a compound such as $Mg_3MoN_4$ would not be stable in a rocksalt-like structure.

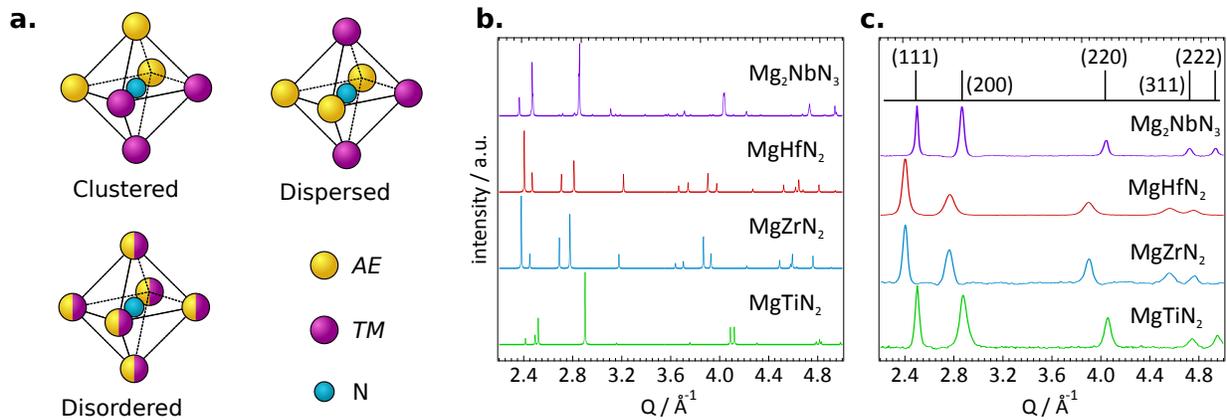

**Figure 1: Structural properties of $Mg_{G-3}TMN_{G-2}$ ($TM$=Ti, Zr, Hf, Nb) materials**. (a) The charge-neutral local structural motifs that serve as building-blocks for $AE^{2+}TM^{4+}N_2$ rocksalt-derived structures. An alternative to the clustered and dispersed cation-ordered motifs is a substitutionally disordered cation lattice. (b) Computed diffraction patterns of the rocksalt-derived ground states for $Mg_{G-3}TMN_{G-2}$ with an ordered cation sublattice (c) Synchrotron X-ray diffraction data exhibit peaks that can be indexed to simple rocksalt (peak positions shown in black for *a*=4.46Å to match Mg$_2$NbN$_3$), indicating the presence of substitutional disorder on the cation sublattice.

In order to confirm these hypothetical considerations about the structures in the Mg$_{G-3}$*TM*N$_{G-2}$ family, we performed first-principles structure searches, using the kinetically limited minimization approach,[7] and following a broader identification of underexplored ternary nitride chemical spaces.[20] The results from our search recovers the experimentally known crystal structures of Sr*TM*$^{4+}$N$_2$ and Ba*TM*$^{4+}$N$_2$[13] and predicts the same structure types that were identified for Mg*TM*$^{4+}$N$_2$ in a recent computational study out of a limited set of prototype structures.[14] For the Mg*TM*$^{4+}$N$_2$ compounds, we obtained cation ordered rocksalt-derived structures, i.e., the layered α-NaFeO$_2$ structure with the space group (SG) number 166 for MgTiN$_2$ and the LiFeO$_2$ structure (SG 141) for MgZrN$_2$ and MgHfN$_2$. The α-NaFeO$_2$ structure is also found as the ground state for all Ca*TM*N$_2$ (*TM* = Ti, Zr, Hf) nitrides. For Mg$_2$NbN$_3$, we find an interesting competition between 5-fold and 6-fold coordination. The lowest energy is obtained for a 5-fold coordinated structure (SG 15) similar to hexagonal boron nitride, however, a rocksalt-derived structure (SG 12) is only 3.5 meV/atom higher in energy. **Figure 1b** shows computed X-ray diffraction patterns for the rocksalt-derived structures and both crystal structure schematics (Figure S1) and cif files are provided in the supplementary information. In summary, for each system the lowest energy rocksalt structure exhibits local octahehedral motifs as expected by the qualitative arguments above; also, our calculations show only a small energy penalty for adopting different motifs (Table S1), meaning these compounds can likely be disordered under non-equilibrium synthesis.

To validate these predictions, we synthesized both stoichiometric (x=0) and Mg-rich (x≈0.5) Mg$_{G-3+x}$*TM*$_{1-x}$N$_{G-2}$ (*TM*=Ti, Zr, Nb, Hf) thin films by combinatorial sputtering on glass substrates. X-ray diffraction data for stoichiometric samples are shown in **Figure 1c**. For each *TM* the peaks can be indexed to a NaCl-type rocksalt structure (SG 225), shown at the top with $a$=4.46Å for reference. Diffraction data from Mg-rich films (Fig. S2) are similar, but with a different preferential orientation. Peak widths (Table S2), indicate crystalline coherence lengths on the order of 10 nm, consistent with observations from transmission electron microscopy (TEM) images (Fig. S3). TEM energy dispersive X-ray spectroscopy measurements also support the absence of amorphous $Mg_3N_2$ or *TM*-N phases in the films. The experimentally determined lattice parameters are in good agreement with calculated structures when the cell volume is reduced to simple rocksalt (Table S2). The lack of additional reflections, which would be present for the rocksalt-derived ground state structures (Figure 1b), suggest substitutional disorder on the cation lattice. This observation is not surprising given the similar ionic radii of Mg (0.86 Å) and the *TM*s (0.75-0.85 Å)[13,21], the prevalence of cation disorder in closed-shell oxide rocksalts[22], and the similar formation energy of structures with different local motifs (Table S1).

Next, we report on optoelectronic properties of the $Mg_{G-3+x}TM_{1-x}N_{G-2}$ thin films. The stoichiometric films prepared for this study had conductivities >1S cm$^{-1}$ and did not exhibit a clearly-defined absorption onset, presumably due to degenerate carrier densities. Thus, following the strategy to prepare non-degenerate Zn-rich $ZnSnN_2$ ternary wurtzite semiconductors by creating oxygen-impurity compensating defect complexes[10], we synthesized Mg-rich compounds, namely, $Mg_{1.54}Ti_{0.46}N_2$, $Mg_{1.54}Zr_{0.46}N_2$, $Mg_{1.52}Hf_{0.48}N_2$, and $Mg_{2.46}Nb_{0.54}N_3$. Increasing Mg content resulted in an increase in measured oxygen, from ca. 3 atomic % (stoichiometric) to ca. 7 atomic % (Mg-rich). However, since this increase lags the Mg increase, we expect charge compensation can still occur. Variable-temperature resistivity measurements from Mg-rich films are presented as an Arrhenius plot in **Figure 2a**. The resistivity exponentially increases with decreasing temperature, which is usually associated with thermally activated charge transport in a semiconductor. For each of the four materials, the resistivities are significantly lower at stoichiometric compositions, confirming that extrinsic carriers are compensated in Mg-rich films. Negative Seebeck voltages measured from MgZrN$_2$ confirm n-type transport, likely from uncompensated oxygen acting as electron donors. As presented in **Figure 2b**, each Mg-rich composition exhibits an absorption onset in the visible range (1.8-2.1 eV), with slight shifts in energy ($E_{Zr} \approx E_{Hf} < E_{Ti} < E_{Nb}$) for a given absorptivity.

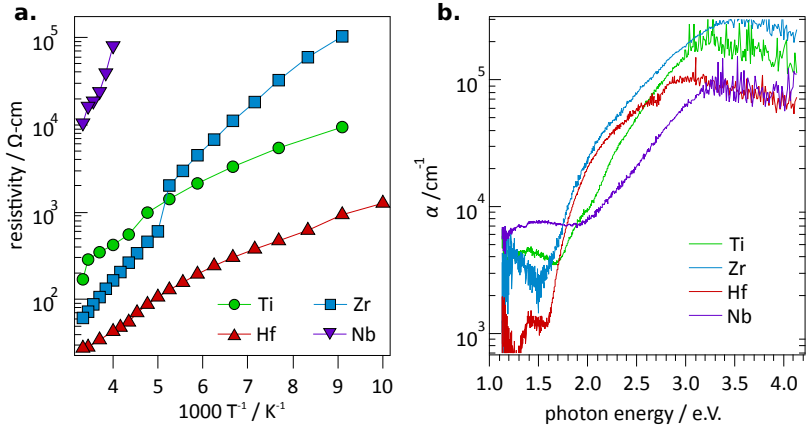

**Figure 2: Optoelectronic properties of Mg-rich Mg$_{G-3+x}$$TM_{1-x}$N$_{G-2}$ materials.** (a) Semiconducting behavior is supported by an exponential increase in resitivity with decreasing temperature. (b) Experimental optical absorption spectra exhibit absorption onsets in the visible range.

To support these experimental measurements, we computed the properties of the Mg$_{G-3}$$TM$N$_{G-2}$ materials with electronic structure calculations using many-body perturbation theory in the GW approximation.[23] The calculation results are summarized in **Table 1**. Electronic band gaps of the rocksalt-derived structures are in the range between 1 and 2 eV, but the optical absorption, shown in **Figure 3a** for MgZrN$_2$, exhibits a slow onset, nearly 1 eV above the band gap, due to the indirect/forbidden nature of the optical transition. Similar data overlayed with experimentel traces are presented for the other Mg$_{G-3}$*TM*N$_{G-2}$ compounds in Figure S4. The effective electron and hole masses derived from density of states calculations are quite low, between 0.6 and 1.9 $m_e$ (Table 1). Notably, we obtain very large dielectic constants between 30-80 $\varepsilon_0$ (vacuum permittivity). This is likely because the $d^0$ configuration of the *TM* cations allows for large cation displacements at minute energy cost, thereby leading to a large ionic contribution to the dielectric constant.[22]

**Table 1:** Calculated properties: Electronic band gap types, electronic band gaps, optical absorption onsets ($E_g^{opt}$, defined as energy when $\alpha = 10^3$ cm$^{-1}$), density of states effective masses, and the static dielectric constants.

|  | Gap type | $E_g$ (eV) | $E_g^{opt}$ (eV) | $m^*_e/m_0$ | $m^*_h/m_0$ | $\varepsilon/\varepsilon_0$ |
|---|---|---|---|---|---|---|
| MgTiN$_2$ | Indirect | 0.91 | 2.1 | 1.4 | 1.4 | 80 |
| MgZrN$_2$ | Direct forbidden | 1.47 | 2.5 | 0.6 | 1.6 | 39 |
| MgHfN$_2$ | Direct forbidden | 1.79 | 2.9 | 0.6 | 1.5 | 32 |
| Mg$_2$NbN$_3$ | Indirect | 1.84 | 2.7 | 0.6 | 1.9 | 68 |

To understand the effect of Mg-*TM* disorder on material properties, we performed Monte-Carlo supercell calculations, using the effective temperature ($T_{eff}$) concept to quantify disorder.[24] We considered both a moderate level of disorder ($T_{eff} \approx$ 2000K), common for oxides and nitrides[24,25], and strong disorder ($T_{eff} \approx$ 10000K). The detailed energetics of the disordered structures are given in Table S3. As seen in Figure 3a for MgZrN$_2$, disorder alleviates the selection rules, thereby bringing the absorption onset close to the band gap energy. However, band gaps and effective masses are only marginally affected by moderate and even strong disorder (Table 1 and Table S4, $\Delta E_g$=0.1-0.3 eV and $\Delta m^*_e$=0.1-0.7 $m_e$), despite considerable deviations from the ideal octet-rule conserving motifs in disordered Mg$_{G-3}$$TM$N$_{G-2}$ (e.g., N-Mg$_3$Ti$_3$ and N-Mg$_4$Nb$_2$). We also note that the disorder stabilizes the rocksalt-like structure of Mg$_2$NbN$_3$ over the distorted hexagonal structure of the ordered ground state (*cf.* Table S3), explaining why a rocksalt structure is observed in experiment.

To further examine the role of structural disorder on electronic properties of Mg$_{G-3}$$TM$N$_{G-2}$ compounds we performed inverse participation ratio (IPR) calculations[26,27] on GS, moderately- and strongly-disordered MgZrN$_2$. **Figure 3b** shows that the IPR of the ground-state structure is around 2 near the band edges, indicating about half the atoms contribute to these states, as is the case for many compound semiconductors (e.g. GaAs). In the disordered materials, the magnitude of the IPR in the vicinity of the band edges is hardly affected, and no localized mid-gap states are observed. Thus, these IPR data indicate that MgZrN$_2$ and related semiconductors are electronically highly tolerant to disorder, even for the "worst-case" scenario of $T_{eff}$= 10,000 K. The results are striking when compared to tetrahedrally coordinated ternary nitrides, where already for $T_{eff}$= 2,000 K the IPR increased by over two times at the VB edge with a concomitant bandgap reduction by one third.[19] We attribute this electronic tolerance towards cation disorder to the exceptionally strong dielectric response in the rocksalt structure, where the potentially adverse effects of breaking of the local octet rule are largely screened out. In a broader context, these findings suggest new opportunities for materials design outside of the tetrahedrally bonded semiconductors.[28,29]

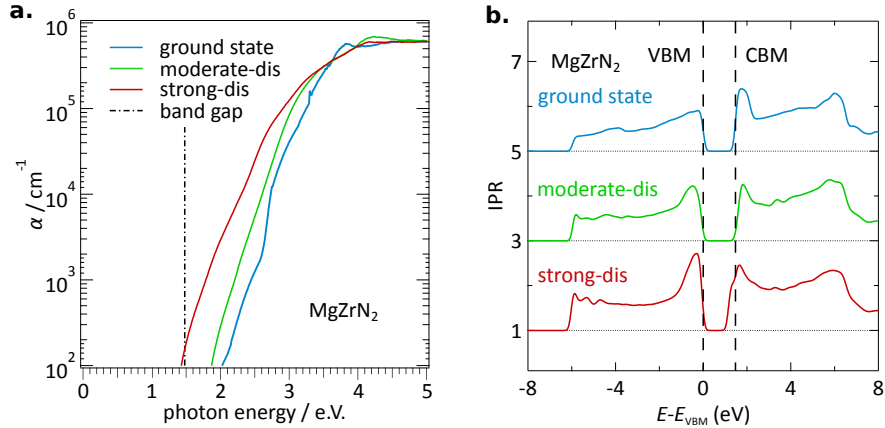

**Figure 3: Calculated properties of MgZrN$_2$.** (a) Absorptivity for MgZrN$_2$ with the rocksalt-derived ground state structure, and disordered structures with moderate (moderate-dis) and strong (strong-dis) disorder. The electronic band gap of the ordered ground state structure (LiFeO$_2$) is shown by a vertical line. The absorption edge shifts towards the band gap with increasing disorder. (b) Inverse participation ratio (IPR) data for MgZrN$_2$ show that carrier localization and band gap change negligibly with disorder.

The calculated structural and optoelectronic properties of the Mg$_{G-3}$TMN$_{G-2}$ rocksalt materials are summarized in **Figure 4**, as inspired by "band gap engineering"[30] in zincblende and wurtzite main-group semiconductor materials. Figure 4a shows band gap vs. effective lattice parameter for the rocksalt-derived crystal structures of Mg$_{G-3}$TMN$_{G-2}$, in comparison with related nitrides.[31,32] Also included is Mg$_2$TaN$_3$, which was not investigated experimentally in this study, but is predicted to assume the same rocksalt-derived structure as Mg$_2$NbN$_3$ (Table S1). As shown in Figure 4a, the chemical space of group-III transition metal nitride rocksalts and heterovalent *AE-TM*-N rocksalts contains several compounds with a band gap range of ca. 1-2 eV. The semiconducting rocksalt nitrides appear to fill the band gap vs. lattice parameter "green gap" in III-N wurtzite compounds. However, the optical absorption onsets are higher in energy than the electronic gap for the cation-ordered structures. Because the absorption onset converges towards the electronic gap at high $T_{eff}$ but the electronic properties are hardly affected (Figure 3), cation disorder engineering might allow for strategically filling the green gap in III-N materials.

As shown in Figure 4a, the ternary rocksalt nitrides have effective cubic lattice parameters between 4.3Å-4.9Å and effective hexagonal lattice parameers of 3.0Å-3.6Å. Since *TM*-N rocksalts with a similar span of lattice parameters are known to form as ternary alloys and superlattices,[33,34] we postulate that similar materials could be made from *AE-TM*-N compounds. The hexagonal projection of Mg$_{G-3}$*TM*N$_{G-2}$ lattice parameters also fall within the range of InN-GaN-AlN *a*-lattice parameters (Figure 4a). Epitaxial growth of (111) ScN has been demonstrated on hexagonal GaN and Al$_2$O$_3$, as well as highly mismatched cubic Si and MgO substrates (lattice parameters shown as black lines in Figure 4a).[34–36] Preliminary experiments suggest similar epitaxial growth techniques to be effective for the Mg$_{G-3}$*TM*N$_{G-2}$ materials: MgZrN$_2$ grown by sputtering on (111) MgO exhibits highly promising transport measurements, with mobilities approaching 100 cm$^2$ V$^{-1}$s$^{-1}$ as presented in **Table 2**. These are very promising results for sputtered and disoreder materials.

**Table 2:** Transport properties of MgZrN$_2$ grown on MgO substrates. Error is standard deviation from 3 measurements.

| Substrate | $\rho$ ($\Omega$-cm) | $\mu$ (cm$^2$V$^{-1}$s$^{-1}$) | $n$ (cm$^{-3}$) |
|---|---|---|---|
| (111) MgO | 0.0192 | 90 ± 60 | (-5 ± 3)×10$^{18}$ |
| (100) MgO | 0.1118 | 42 ± 9 | (-1.3 ± 0.3)×10$^{20}$ |

Figure 4b shows a color map of calculated optical absorption onsets (defined here by a threshold $\alpha=10^3$ cm$^{-1}$) for Mg$_{G-3}$$TM$N$_{G-2}$ with contour lines showing electron effective masses, both as a function of effective temperature and lattice parameter. The 2D linear interpolation is generated from the calculated stoichiometric values highlighted by white hexagons for ground state, moderately disordered, and strongly disordered structures. Even though disorder does not significantly affect the effective masses or band gaps, absorption onsets decrease in energy with disorder. Thus, the degree of disorder could be utilized to tune the absorption strength to the needs of the application without deteriorating charge transport properties. Since the absorption onset is generally more sensitive to disorder than to lattice parameter, these "disorder-tuned" materials might be integrated with existing materials via alloying. Overall, Figure 4 suggests both composition-based and disorder-based tuning of optoelectronic properties should be possible in Mg$_{G-3}$$TM$N$_{G-2}$, providing an additional degree of freedom not found in binary nitrides or materials with less defect tolerance.

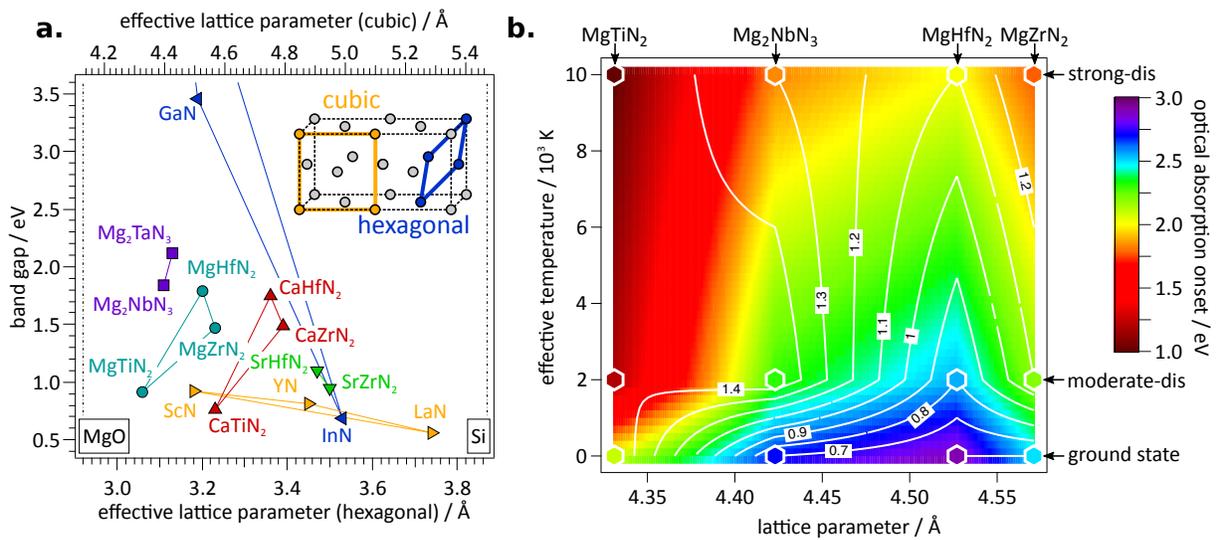

**Figure 4: Tunable properties of Mg$_{G-3}$*TM*N$_{G-2}$ semiconductors**. (a) Band gap vs. effective lattice parameter for Mg$_{G-3}$*TM*N$_{G-2}$, group-III nitrides, and *AE-TM*-N$_2$ materials.[31,32] The inset illustrates the anion sublattice for a rocksalt-derived structure showing how effective lattice parameters are determined. These data suggest lattice compatibility between ternary nitride rocksalt semiconductors and cubic substrates (MgO, Si, black vertical lines) as well as wurzite nitrides. (b) Interpolation of optical absorption onset (color) and electron effective mass (contour lines) between different Mg$_{G-3}$*TM*N$_{G-2}$ lattice parameters and degrees of disorder, as indicated by effective temperature. The calculated values, outlined by white hexagons, are for ground state, moderately disordered (moderate-dis), and strongly disordered (strong-dis) ternary Mg$_{G-3}$*TM*N$_{G-2}$.

  In conclusion, we have introduced a new family of Mg$_{G-3}$*TM*N$_{G-2}$ (*TM* = Ti, Zr, Nb, Hf) semiconducting materials with rocksalt-derived crystal structures, and studied their optoelectronic properties using both sputtered thin films and first-principles calculations. For each material of interest the measured diffraction patterns are consistent with the calculated rocksalt-derived structures, but with significant antisite disorder on the cation sublattice. When the materials are synthesized with Mg-rich stoichiometry, they exhibit semiconducting properties, including absorption onsets in the visible range. Ab-initio calculations suggest that the optical absorption onset can be tuned both by choice of transition metal and the degree of ordering on the cation sublattice, while the band gaps and effective masses exhibit a remarkable tolerance to structural disorder. The potential for integrating these novel rocksalt semiconductors into optoelectronic devices is highlighted by their structural compatibility with existing optoelectronic nitrides and promising transport properties from sputtered epitaxial films.

**Methods**

**Synthesis:** $Mg_xTM_{1-x}N_y$ ($TM$ = Ti, Zr, Nb, Hf) films were deposited on 50.8x50.8x1.1mm³ glass substrates via radio frequency magnetron cosputtering at 30-60 watts from metallic targets (Mg 99.98+%, all $TM$ 99.9+%). Sputter deposition occurred in close proximity to a cryogenic sheath at a pressure of 5 mT under 6 sccm of both 99.999+% Ar and $N_2$ gases in a vacuum chamber with base pressure < $10^{-7}$ torr.[37] Sputter targets were oriented obliquely to the stationary substrates such that a 1-D gradient in metal composition was created.[38] During growth, an orthogal temperature gradient was passively induced by partial thermal contact of the glass substrate to a heated platen. The contiuously varying processing conditions present on the sample surface were discretized into a 44 point grid. The specific points described in this study were selected for growth temperatures of 400°C and $Mg_{G-3+x}TM_{1-x}N_{G-2}$ compositions of x=0 (stoichiometric) and x≈0.5 (Mg-rich). Hall measurements were made on separately prepared samples deposited on (100)- and (111)-oriented MgO substrates, positioned to target the $MgZrN_2$ composition. Process conditions were kept the same, but with nitrogen introduced through an RF cracker operating at 350W.

**Characterization:** X-ray diffraction data presented in the main text were collected at the Stanford Synchrotron Radiation Lightsource on beamline 1-5 at an energy of 15.5 keV on a CCD area detector. Rutheford backscattering (RBS) measurements were performed on a National Electrostatics Corporation 3S-MR10 instrument with a 2MeV alpha particle beam. Compositions from RBS data were determined using the RUMP analysis package.[39] Film thickness was determined by stylus profilometry using a Dektak 8 profilometer. Transmission and reflection spectra were collected on a custom thin film optical spectroscopy system and absorption spectra were determined from these data together with film thickness. Transport measurements from Mg-rich compositions and epitaxial films were measured using a Lakeshore 8425 Hall probe equipped with a 2T superconducting magnet and variable temperature sample environment. Transmission electron microscopy (TEM) was performed using an FEI Talos F200X microscope operating at an accelerating voltage of 200 kV. Energy dispersive x-ray spectroscopy was performed on the TEM in scanning mode. The TEM lamella was prepared via ion beam milling and lift-out using an FEI Helios Nanolab dual beam SEM/FIB.

**Computational:** First principles density functional and many-body perturbation theory calculations were performed with the VASP code, employing the generalized gradient (GGA) and GW approximations, respectively[40,41], with an on-site Coulomb interaction of $U = 3eV$ for the *TM* cations. The ground state structure search was performed using the "kinetically limited minimization" approach, which is unconstrained and does not require prototypical structures from databases.[7] Seed structures are generated from random lattice vectors and atomic positions, subject to geometric constraints to avoid extreme cell shapes, and to observe minimal interatomic distances (2.8 Å for cation-cation and anion-anion pairs, 1.9 Å for cation-anion pairs). New trial structures are generated by the random displacement of one atom between 1.0 and 5.0 Å while maintaining the minimal distances. Trial structures are accepted if the total energy is lowered, and the number of trials equals the number of atoms in the unit cell. For each material, we sampled at least 100 seeds, using 16 atom cells for $AE_1TM_1N_2$ and 12 atom cells for $Mg_2NbN_3$. For the final ground state structures, a symmetry analysis was performed using the FINDSYM software[42], and a Crystallographic Information File (CIF) was generated (available in the supplementary information). GW calculations were performed for these structures as described in Ref.[31] Note that for compounds with the α-NaFeO$_2$ structure (SG 166) we find a slight distortion accompanied by an energy lowering of less than 1 meV/at, technically being described as a SG 13 structure with a larger 8 atom primitive cell and direct-forbidden electronic band gap. Disordered structures were generated through first-principles Monte-Carlo sampling in supercells between 64 and 96 atoms (5 random seeds per each case) using the Metropolis criterion for 2000 and 10000 K as effective temperature[24] for moderate and strong disorder, respectively. In order to calculate the electronic structure and optical absorption for these supercells, we used the single-shot-hybrid plus onsite potential (SSH+V) approach[19,31], with parameters fitted to the GW calculations of the ground states (see Table S5). The density of states effective masses were determined as described in Ref.[43]

**Acknowledgments**

This work was supported by the U.S. Department of Energy (DOE) under Contract No. DE-AC36-08GO28308 with Alliance for Sustainable Energy, LLC, the Manager and Operator of the National Renewable Energy Laboratory. Funding provided by Office of Science (SC), Office of Basic Energy Sciences (BES), as part of the Energy Frontier Research Center "Center for Next Generation of Materials Design: Incorporating Metastability". High-performance computing resources were sponsored by the DOE, Office of Energy Efficiency and Renewable Energy. Use of the SSRL, SLAC National Accelerator Laboratory, was supported by the DOE, SC, BES under contract no. DE-AC02-76SF00515. Work by A. Tamboli and C. Melamed was supported by the DOE, SC, BES, Materials Sciences and Engineering Division. We would like to thank Dr. Suchismita Sarker and Dr. Apurva Mehta for assistance at SLAC BL1-5.

## Author Contributions

S.R.B. made the thin films, designed experiments, and analyzed results. A.H., W.S., and S.L. performed DFT calculations. C.L.M., R.W.R., and AT helped collect synchrotron XRD data. J.M. collected and analyzed TEM data. J.P. collected and analyzed RBS data. S.R.B., A.H., W.S., J.P., W.T., B.G., G.C., S.L., and A.Z. conceived the project. S.R.B., S.L., and A.Z. prepared the manuscript with input from all authors.

**Competing Interests:** The authors declare no competing interests.

# Supporting Information for Ternary Nitride Semiconductors in the Rocksalt Crystal Structure

*Sage R. Bauers[*,1], Aaron Holder[1,2], Wenhao Sun[3,4], Celeste L. Melamed[1,5], Rachel Woods-Robinson[1,3,4], John Mangum[5], John Perkins[1], William Tumas[1], Brian Gorman[5], Adele Tamboli[1], Gerbrand Ceder[3,4], Stephan Lany[1], Andriy Zakutayev[*,1]*

[1]Materials Science Center, National Renewable Energy Laboratory, Golden, Colorado 80401, United States
[2]University of Colorado, Boulder, Colorado 80309, United States
[3]Materials Science Division, Lawrence Berkeley National Laboratory, Berkeley, California 94720, United States
[4]University of California Berkeley, Berkeley, California 94720, United States
[5]Colorado School of Mines, Golden, Colorado 80401, United States
[*]Corresponding authors

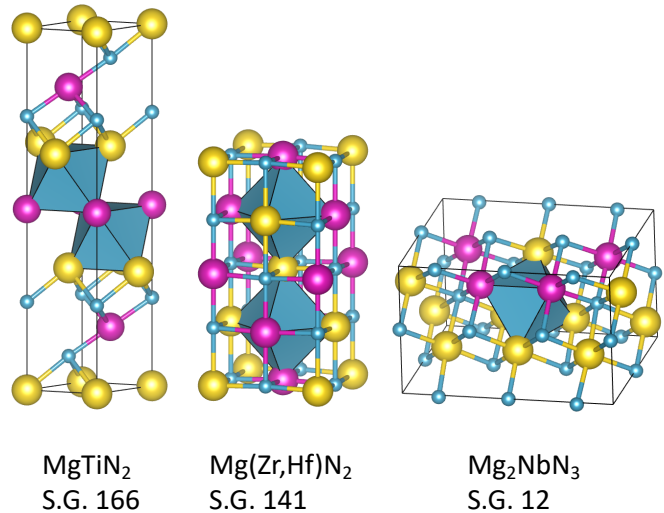

MgTiN$_2$         Mg(Zr,Hf)N$_2$    Mg$_2$NbN$_3$
S.G. 166          S.G. 141          S.G. 12

**Figure S1:** Extended ordered rocksalt-derived structures of Mg$_{G-3}$*TM*N$_{G-2}$ (*TM*=Ti, Zr, Nb, Hf) ternary nitrides, with N shown in cyan, Mg in yellow, and *TM* in magenta. The nitrogen centered octahedra arrange as clustered motifs in S.G. 166 and S.G. 12 and as a dispersed motif in S.G. 141 (S.G.= Space Group).

**Table S1:** Calculated polymorph energies of the Mg$_{G-3}$*TM*N$_{G-2}$ ternary nitrides studied in this work (S.G.= Space Group).

| Compound | S.G. 129 (eV/atom) | S.G. 141 (eV/atom) | S.G. 166 (eV/atom) | S.G. 15 (eV/atom) | S.G. 12 (eV/atom) |
|---|---|---|---|---|---|
| MgTiN$_2$ | 0.356 | 0.033 | 0.000 | | |
| MgZrN$_2$ | 0.470 | 0.000 | 0.016 | | |
| MgHfN$_2$ | 0.482 | 0.000 | 0.018 | | |
| Mg$_2$NbN$_3$ | | | | 0.000 | 0.003 |
| Mg$_2$TaN$_3$ | | | | 0.017 | 0.000 |

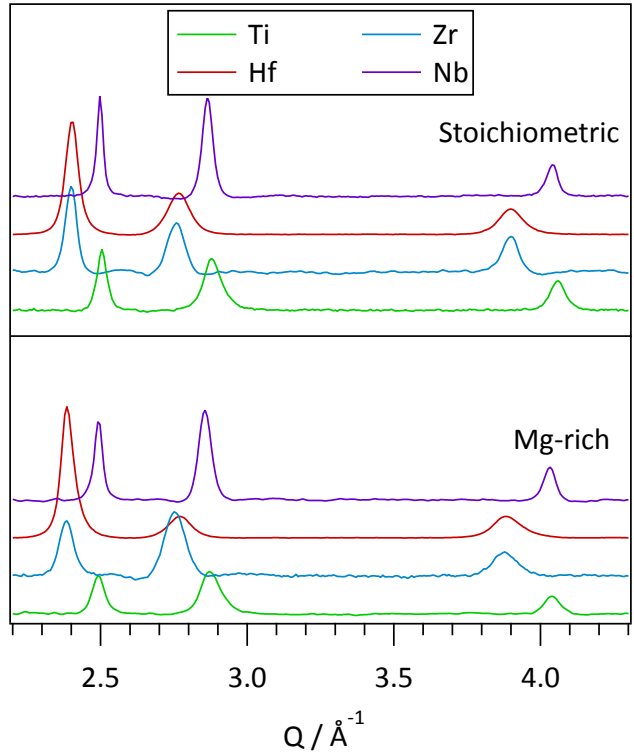

**Figure S2:** Diffraction patterns from stoichiometric (top; x=0) and Mg-rich (bottom; x≈0.5) $Mg_{G-3+x}TM_{1-x}N_{G-2}$ thin films (G=4 for Ti, Zr, Hf and G=5 for Nb). The reflections for each pattern can be indexed to the NaCl (space group 225) structure.

**Table S2:** Lattice parameters, peak widths, and absorption onsets ($\alpha=10^4$ cm$^{-1}$) for experimental (Exp) $Mg_{G-3+x}TM_{1-x}N_{G-2}$ thin films (G=4 for Ti, Zr, Hf and G=5 for Nb). Absorption values are for Mg-rich compositions (x≈0.5). Also displayed are calculated values for the predicted volume-based lattice parameters of the rocksalt-derived ground-state (GS) $Mg_{G-3}TMN_{G-2}$ structures, and the calculated absorption onset ($\alpha=10^3$ cm$^{-1}$) of $Mg_{G-3}TMN_{G-2}$ in ground-state (GS, $T_{eff}$ = 0 K), and with moderate- (mod-D, $T_{eff}$ = 2000 K), and strong-disorder (str-D, $T_{eff}$ = 10000 K).

| Compound | (111) FWHM (Å$^{-1}$) | Rocksalt lattice parameter (Å) | | Energy at $\alpha=10^4$ cm$^{-1}$ (Exp) or $\alpha=10^3$ cm$^{-1}$ (calculated) (eV) | | | |
|---|---|---|---|---|---|---|---|
| | Exp | Exp | GS | Exp | GS | Mod-D | Str-D |
| MgTiN$_2$ | 0.042 | 4.44 | 4.33 | 2.0 | 2.1 | 1.2 | 1.0 |
| MgZrN$_2$ | 0.065 | 4.63 | 4.57 | 1.8 | 2.5 | 2.2 | 1.8 |
| MgHfN$_2$ | 0.053 | 4.62 | 4.53 | 1.9 | 2.9 | 2.6 | 2.0 |
| Mg$_2$NbN$_3$ (rocksalt) | 0.025 | 4.46 | 4.42 | 2.1 | 2.7 | 2.3 | 1.8 |

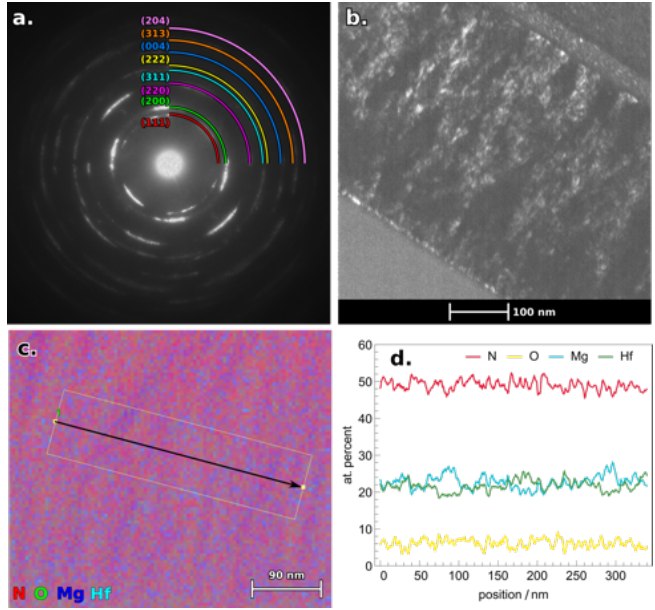

**Figure S3:** TEM characterization of a $MgHfN_2$ lamella. **a**. SAED pattern exhibits reflections that are indexed to the NaCl structure. **b**. Dark field image highlights individual crystallites that are ~10nm in size. **c**. STEM-EDS map appears predominantly homogeneous, but with faint curtaining suggesting small fluctuations. **d**. Quantification of EDS map agrees with composition measured via Rutheford Backscattering and shows slight anti-correlated modulation in Mg and Hf compositions.

**Table S3:** Structural energetics of disordered Mg-*TM*-N structures from Monte-Carlo simulations. Energies above the ground state for moderate ($T_{eff}$ = 2000 K), strong ($T_{eff}$ = 10000 K), and fully random ($T_{eff} = \infty$) disorder.

| Compound | moderate (meV/at) | strong (meV/at) | random (meV/at) |
|---|---|---|---|
| $MgTiN_2$ | 18±4 | 30±6 | 52±8 |
| $MgZrN_2$ | 16±3 | 22±8 | 32±7 |
| $MgHfN_2$ | 20±3 | 28±9 | 45±11 |
| $Mg_2NbN_3$ (hex) | 24±7 | 53±14 | 123±34 |
| $Mg_2NbN_3$ (rocksalt) | 13±3 | 35±9 | 45±6 |

**Table S4:** Electronic properties of disordered Mg-*TM*-N in rocksalt structures.

| compound | moderate | | | strong | | |
|---|---|---|---|---|---|---|
| | $E_g$ (eV) | $m^*_h/m_0$ | $m^*_e/m_0$ | $E_g$ (eV) | $m^*_h/m_0$ | $m^*_e/m_0$ |
| $MgTiN_2$ | 0.92 | 1.4 | 1.4 | 0.81 | 1.5 | 1.5 |
| $MgZrN_2$ | 1.67 | 2.1 | 1.1 | 1.39 | 2.4 | 1.3 |
| $MgHfN_2$ | 1.92 | 2.0 | 0.8 | 1.79 | 2.1 | 1.1 |
| $Mg_2NbN_3$ | 2.14 | 3.6 | 1.5 | 1.66 | 3.7 | 1.3 |

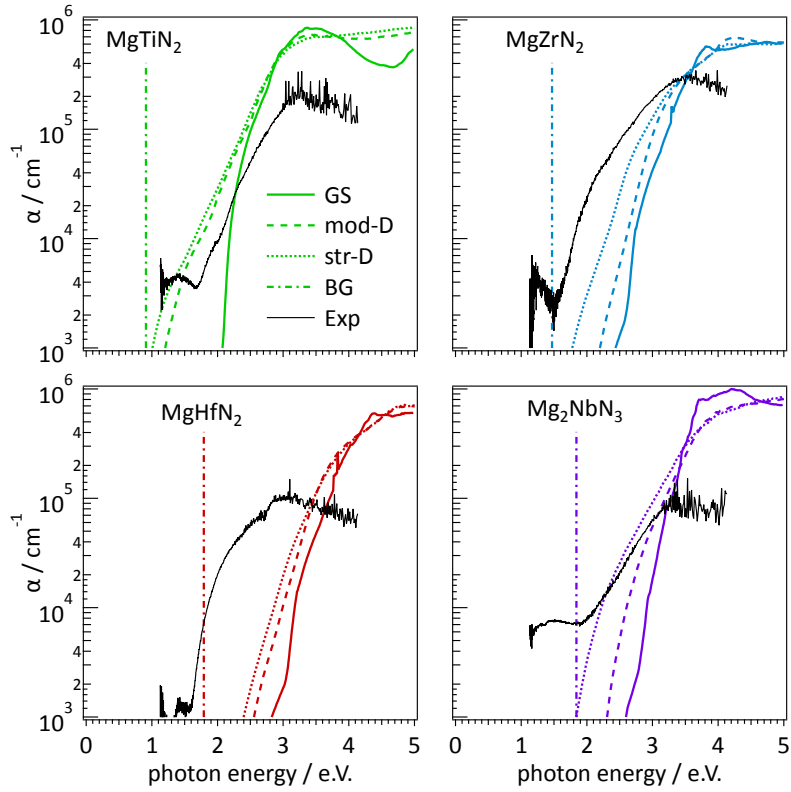

**Figure S4:** Calculated absorption spectra for $Mg_{G-3}TMN_{G-2}$ with the rocksalt-derived ground state (GS) structure, and disordered structures with moderate- (mod-D) and strong- (str-D) disorder. The band gap of the ordered GS structure is shown by a vertical line. Overlayed are measured spectra for Mg-rich compositions (Exp).

**Table S5:** Parameters for non-self-consistent SSH+$V$ calculations. The values for the Fock exchange ($\alpha$) and onsite potential ($V$) were fitted to GW calculations of the ordered ground states to reproduce the GW band structure within about 0.02 eV at $E_g$ and about 0.1 eV at $2\times E_g$. The onsite-potential acts on the $TM$-$d$ orbitals only. For $Mg_2NbN_3$, both the hexagonal lattice (hl) and rock-salt derived (rs) structures were used.

|       | $MgTiN_2$ | $MgZrN_2$ | $MgHfN_2$ | $Mg_2NbN_3$-hl | $Mg_2NbN_3$-rs |
|-------|-----------|-----------|-----------|----------------|----------------|
| $\alpha$ (%) | 10.2 | 11.9 | 14.3 | 10.6 | 13.1 |
| $V$ (eV) | 0.21 | 0.65 | 0.27 | 1.17 | 1.30 |